\documentclass{article}
\usepackage{spconf,amsmath,amssymb,amsfonts,graphicx,cite}
\usepackage{booktabs}
\usepackage{algorithm}
\usepackage{algorithmic}
\usepackage[hidelinks]{hyperref}

\title{DUAL-HYPERGRAPH INDEXING: BRIDGING KNOWLEDGE ISLANDS FOR MULTI-HOP REASONING IN RETRIEVAL-AUGMENTED GENERATION}

\name{Qi Sun\textsuperscript{1*}, Caibo Li\textsuperscript{2}, Yijia Zhang\textsuperscript{2}, Qiang Li\textsuperscript{3}, Yu Guo\textsuperscript{1\textdagger}\thanks{\textdagger\ Corresponding author: yu.guo@xjtu.edu.cn}}
\address{\textsuperscript{1}School of Software Engineering, Xi'an Jiaotong University, Xi'an, Shaanxi, China\\
\textsuperscript{2}State Key Laboratory of Human-Machine Hybrid Augmented Intelligence,\\ and Institute of Artificial Intelligence and Robotics, Xi'an Jiaotong University, Xi'an, Shaanxi, China\\
\textsuperscript{3}EHV Power Transmission Company of China Southern Power Grid Co., Ltd.
}

\begin{document}
\ninept 
\maketitle

\begin{abstract}

While hypergraph-based Retrieval-Augmented Generation (RAG) effectively captures higher-order multi-entity correlations, existing paradigms treat extracted hyperedges as isolated factual assertions. This structural fragmentation engenders rigid ``knowledge islands'' that bottleneck multi-hop causal inference, temporal tracking, and narrative synthesis. To systematically address these challenges, we introduce Dual-Hypergraph Indexing (DHI), a hierarchical representation framework that elevates discrete facts into structured analytical insights. DHI couples a foundational entity-relation factual hypergraph ($H_K$) with an elevated deep-insight hypergraph ($H_D$) via a dual-pathway aggregation algorithm. Specifically, DHI employs: (1) importance-driven hub aggregation via 5-metric topological profiling and adaptive thresholding to capture spatial semantic clusters; and (2) temporal chunk-chain progressive aggregation via sliding-window greedy exploration to track chronological evolutions. Across five benchmarks, DHI achieves state-of-the-art performance, boosting logical coherence by +1.53 on the multidisciplinary Mix benchmark and scoring 85.78\% on complex medical pathology reasoning tasks. DHI provides a robust architecture for next-generation multi-hop RAG.
\end{abstract}


\section{INTRODUCTION}
\label{sec:intro}

The advent of Large Language Models (LLMs) has fundamentally transformed natural language processing, enabling unprecedented capabilities in complex reasoning. However, their parametric memory remains inherently static and susceptible to factual hallucinations, particularly when navigating specialized, proprietary, or rapidly evolving domains \cite{ji2023survey}. Non-parametric retrieval augmentation---commonly known as Retrieval-Augmented Generation (RAG)---curtails these hallucinations by grounding the generative space in dynamically retrieved, verifiable external evidence \cite{lewis2020retrieval, izacard2023atlas}.

As downstream user queries transition from localized, single-fact retrieval toward multifaceted deductive reasoning and multi-hop synthesis, knowledge representation architectures must evolve. Initial paradigms relying on unstructured text chunk embeddings frequently fail to capture cross-document relational semantics \cite{chen2024benchmarking}. To address this, structured graph topographies, such as GraphRAG \cite{edge2024local} and LightRAG \cite{guo2025lightrag}, explicitly construct entity-relation knowledge graphs to enable macro-level community summaries. More recently, hypergraph-driven architectures have demonstrated distinct topological advantages. By natively modeling complex, non-pairwise interactions among multiple entities ($n \ge 3$) as unified hyperedges, hypergraph RAG prevents the severe semantic distortion caused by binary edge decomposition \cite{feng2019hypergraph}.

Despite these topological advancements, contemporary hypergraph RAG architectures predominantly exhibit a critical structural limitation we formally term the ``Knowledge Island Problem''. Specifically, existing pipelines typically extract multi-entity hyperedges independently from localized text chunks \cite{hu2026cog, feng2026hyper}, inheriting the structural fragmentation common in chunk-based retrieval \cite{edge2024local, jin2024graph} and treating each hyperedge as a static, isolated semantic silo. Consequently, these architectures lack explicit mathematical pathways to synthesize broader thematic patterns, global causal dependencies, and longitudinal chronological progressions spanning multiple independent hyperedges. During multi-hop retrieval, standard retrievers supply a disconnected subset of assertions \cite{karpukhin2020dense}. Deprived of pre-computed relational bridges, the generator is forced to infer cross-passage dependencies on the fly---a process highly vulnerable to attention dispersion, cognitive overload, and cascading hallucinations \cite{trivedi2023interleaving, shi2024replug}.

To transcend these limitations, we propose Dual-Hypergraph Indexing (DHI), a novel hierarchical representation framework driven by a robust dual-pathway knowledge aggregation mechanism. DHI fundamentally decouples the indexing structure into a meticulously coordinated two-tiered hierarchy: a foundational factual hypergraph $H_K$ capturing granular atomic assertions, and an elevated deep insight hypergraph $H_D$ modeling synthesized conceptual schemas.

DHI's core computational engine executes two orthogonal aggregation pathways:
\begin{itemize}
    \item \textbf{Importance-Driven Hub Aggregation (Path A):} Captures spatial topological significance. We evaluate a robust 5-dimensional structural profiling vector. Applying P90-normalization and an adaptive elbow cutoff, DHI identifies and synthesizes systemic cross-context observations centered around core domain hubs.
    \item \textbf{Temporal Chunk-Chain Progressive Aggregation (Path B):} Captures longitudinal narrative evolutions and sequential causality. By tracking source-chunk chronological footprints within a constrained sliding window, DHI perfectly bridges sequential causalities spanning vast document distances.
\end{itemize}
By seamlessly bridging disconnected knowledge islands, DHI achieves state-of-the-art logical coherence (85.47 on multidisciplinary benchmarks) while significantly reducing context tokens, establishing a robust blueprint for reasoning-intensive Graph RAG.

\section{RELATION TO PRIOR WORK}
\label{sec:prior}

\subsection{Chunk-based and Graph-based RAG Architectures}
Standard RAG systems (e.g., RETRO \cite{borgeaud2022improving}, NaiveRAG) perform dense vector similarity matching over fixed-length text chunks. While efficient, chunking inherently severs relational continuity across paragraph boundaries \cite{jin2024graph}. To resolve this, GraphRAG \cite{edge2024local} pioneered automated entity-relation graphs paired with Leiden community clustering to generate hierarchical summaries. LightRAG \cite{guo2025lightrag} further optimized this via a dual-level entity-relation scheme, while HippoRAG \cite{gutierrez2024hipporag} introduced neurobiologically inspired memory integration. Moreover, active RAG frameworks \cite{jiang2023active} have explored multi-turn retrieval paths. Nevertheless, traditional pairwise graphs fundamentally fail to represent high-order interactions involving three or more entities simultaneously, leading to unavoidable structural loss \cite{fatemi2019knowledge}.

\subsection{Hypergraph Representation in Information Retrieval}
Hypergraph Neural Networks \cite{feng2019hypergraph} mathematically validate that hyperedges preserve non-decomposable group interactions. Subsequent works further introduced hypergraph attention mechanisms to dynamically capture high-order correlation structures. Hyper-RAG \cite{feng2026hyper} proposed localized hypergraph modeling, while Cog-RAG \cite{hu2026cog} introduced cognitive-inspired theme alignment.

Despite these advances, existing hypergraph RAG approaches almost universally treat extracted hyperedges as static structural entities. They fail to distill cross-hyperedge causal trajectories or longitudinal temporal developments \cite{sun2024think}. DHI departs from these paradigms by formalizing a dual-hypergraph architecture that explicitly synthesizes higher-order insights via synergistic spatial and chronological exploration, directly neutralizing the ``Knowledge Island Problem''.

\section{PROPOSED METHODOLOGY}
\label{sec:method}

\subsection{System Architecture and Formal Definition}
Given a raw textual corpus partitioned into sequentially ordered, overlapping chunks $\mathcal{C} = \{c_1, \dots, c_M\}$, DHI constructs a hierarchically coupled dual-hypergraph framework, defined as $(H_K, H_D)$:
\begin{itemize}
    \item \textbf{Factual Hypergraph $H_K = (V_K, E_K)$:} The foundational layer. Vertices $v \in V_K$ denote named entities. Hyperedges $e \in E_K$ represent simple pairwise ($|e| = 2$) or complex higher-order ($|e| \ge 3$) relations. Every higher-order hyperedge $e$ encapsulates a specific relation summary and an atomic direct insight $\iota(e)$, generated during the initial extraction phase.
    \item \textbf{Deep Insight Hypergraph $H_D = (V_D, E_D)$:} The synthesis layer. Vertices $u \in V_D$ establish a strict bijective mapping with higher-order hyperedges of $H_K$, formulated as $V_D = \{\pi(e) \mid e \in E_K, |e| \ge 3\}$. Each deep insight hyperedge $\varepsilon \in E_D$ binds a targeted subset of these insight vertices, enriched with a synthesized narrative insight $\mathcal{I}(\varepsilon)$.
\end{itemize}

\begin{figure*}[t]
\centering
\includegraphics[width=0.88\textwidth]{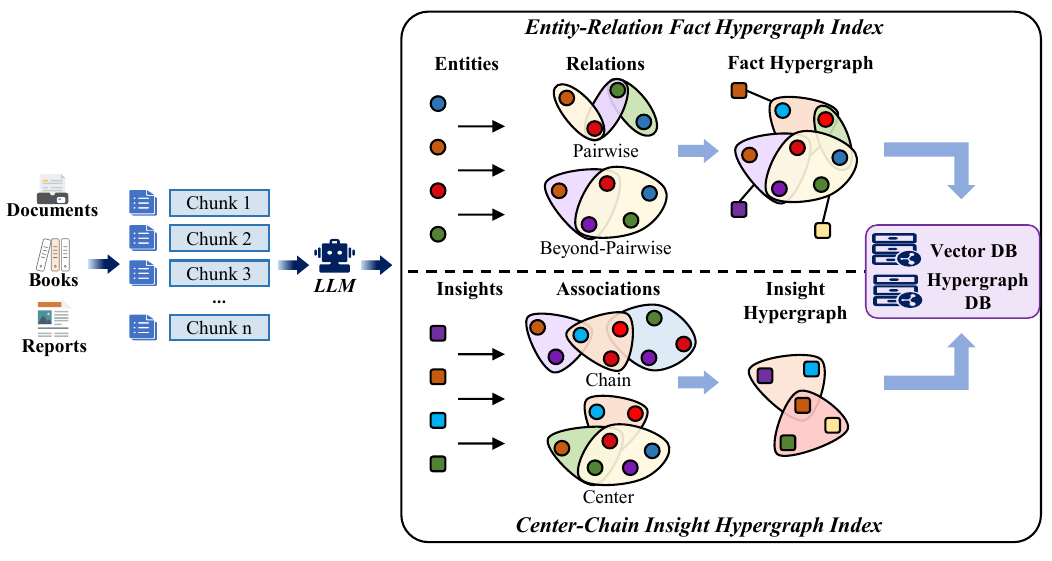}

\caption{The end-to-end operational workflow of the Dual-Hypergraph Indexing (DHI) framework. Source documents are iteratively segmented into sequential sliding chunks. An LLM extracts multi-entity hyperedges equipped with atomic direct insights to construct the foundational factual hypergraph $\mathcal{H}_K$ (bottom). Dual pathways—topological hub aggregation (Path A) and temporal chunk-chain progressive aggregation (Path B)—synthesize these assertions into high-order semantic correlations, populating the deep insight hypergraph $\mathcal{H}_D$ (top). Bidirectional mapping ensures rigorous cross-layer provenance, mitigating hallucinations during context assembly.}
\label{fig:framework}

\end{figure*}

\subsection{The Knowledge Island Formulation}
The necessity of $H_D$ arises from spatial isolation within $H_K$. For a multi-hop query $q$ requiring logical traversal from $v_a$ to $v_c$ via intermediary $v_b$, standard retrieval yields disjoint hyperedges $e_1 = \{v_a, v_b, \dots\}$ and $e_2 = \{v_b, v_c, \dots\}$. The semantic void between $\iota(e_1)$ and $\iota(e_2)$ constitutes the knowledge island boundary. DHI actively computes topological bridges $\varepsilon = \{\pi(e_1), \pi(e_2)\}$ inside $H_D$ prior to retrieval.

\subsection{Dual-Path Insight Aggregation Algorithm}
DHI executes two orthogonal aggregation pathways (Algorithm~\ref{alg:dhi}) targeting distinct axes of information distribution.

\begin{algorithm}[t]
\caption{Dual-Path Dual-Hypergraph Indexing}
\label{alg:dhi}
\begin{algorithmic}[1]
\REQUIRE Sequential Corpus $\mathcal{C}$, Sliding window limit $W = 10$.
\ENSURE Factual Hypergraph $H_K$, Deep Insight Hypergraph $H_D$.
\STATE Parse $\mathcal{C}$ via LLM to extract $V_K$, $E_K$, and direct atomic insights $\iota(e)$.
\STATE Initialize $H_K \leftarrow (V_K, E_K)$.
\STATE Initialize $H_D \leftarrow (V_D = \{\pi(e) \mid e \in E_K, |e| \ge 3\}, E_D = \emptyset)$.
\STATE \textbf{/* Path A: Importance-Driven Hub Aggregation */}
\FOR{each unique entity $v \in V_K$}
    \STATE Compute $\mathbf{m}(v)$ via Eq.~(\ref{eq:profiling}) and robust $S(v)$ via Eqs.~(\ref{eq:norm})--(\ref{eq:score}).
\ENDFOR
\STATE $V_{\text{hub}} \leftarrow \text{ElbowCutoff}(\text{SortDescending}(V_K, S))$.
\FOR{each identified hub entity $v^* \in V_{\text{hub}}$}
    \STATE $\mathcal{I}_{\text{hub}} \leftarrow \text{LLM\_Synthesize\_Hub}(v^*, E(v^*))$.
    \STATE $E_D \leftarrow E_D \cup \{(\{\pi(e) \mid e \in E(v^*)\}, \mathcal{I}_{\text{hub}})\}$.
\ENDFOR
\STATE \textbf{/* Path B: Temporal Chunk-Chain Aggregation */}
\STATE $\text{Chains} \leftarrow \text{GreedyBranchingSearch}(E_K, W)$ using Eq.~(\ref{eq:chain_cond}).
\STATE $\text{Chains}_{\text{pruned}} \leftarrow \text{FilterAndDeduplicate}(\text{Chains}, L_{\min} = 3)$.
\FOR{each valid temporal chain $\mathcal{X} \in \text{Chains}_{\text{pruned}}$}
    \STATE $\mathcal{I}_{\text{chain}} \leftarrow \text{LLM\_Synthesize\_Temporal}(\mathcal{X})$.
    \STATE $E_D \leftarrow E_D \cup \{(\{\pi(e) \mid e \in \mathcal{X}\}, \mathcal{I}_{\text{chain}})\}$.
\ENDFOR
\STATE Construct dense vector database indices for $H_K$ and $H_D$.
\RETURN $H_K, H_D$
\end{algorithmic}
\end{algorithm}

\subsubsection{Path A: Importance-Driven Hub Entity Aggregation}
Path A identifies authoritative semantic hubs. Because vertex degree alone is structurally insufficient for evaluating prominence within hypergraphs \cite{benson2019three}, we formulate a 5-dimensional structural profiling vector for candidate entities $v \in V_K$:
\begin{equation}
\label{eq:profiling}
\mathbf{m}(v) = \big[ d(v), |E(v)|, \bar{s}(v), w(v), |N(v)| \big]^T
\end{equation}
Here, $d(v)$ is vertex degree, $|E(v)|$ signifies incident hyperedge count, $\bar{s}(v) = |E(v)|^{-1} \sum_{e \in E(v)} |e|$ represents average hyperedge scale, $w(v)$ is cumulative semantic weight, and $|N(v)|$ tracks distinct neighbor coverage.

To prevent extreme high-degree generic outliers from statistically obfuscating the true semantic backbone, we implement a P90-robust normalization strategy:
\begin{equation}
\label{eq:norm}
\tilde{m}_k(v) = \frac{\min(m_k(v), P_{90}(m_k)) - m_k^{\min}}{P_{90}(m_k) - m_k^{\min}}
\end{equation}
where $k \in \{1, \dots, 5\}$, $m_k^{\min} = \min_u m_k(u)$, and $P_{90}$ is the 90th percentile. The composite importance score is the uniform mean:
\begin{equation}
\label{eq:score}
S(v) = \frac{1}{5} \sum_{k=1}^5 \tilde{m}_k(v)
\end{equation}

Entities are sorted in descending order of $S(v)$. We autonomously determine the optimal hub inclusion boundary index $k^*$ via maximum first-order difference, bounded by a 20\% system safety cap to prevent insight dilution:
\begin{equation}
\label{eq:elbow}
k^* = \arg\max_k \big(S(v_k) - S(v_{k+1})\big), \quad \text{s.t. } k \le 0.20 \cdot |V_K|
\end{equation}
For each identified hub $v^*$, an LLM synthesizes its entire incident hyperedge set $E(v^*)$ into a systemic hub insight $\mathcal{I}_{\text{hub}}(v^*)$. The insight hyperedge $\varepsilon_{\text{hub}} = \{\pi(e) \mid e \in E(v^*)\}$ is instantiated within $H_D$.

\subsubsection{Path B: Temporal Chunk-Chain Progressive Aggregation}
While Path A extracts static radial clusters, Path B discovers longitudinal narrative chains and chronologies. Let $\tau(c) \in \mathbb{N}^+$ denote the absolute sequential position of chunk $c$. For any hyperedge $e$, its temporal footprint is $\tau(e) = \{\tau(c) \mid c \in \text{source}(e)\}$.

We initialize potential analytical chains and iteratively expand them via a greedy branching algorithm within a sliding window $W$. A candidate hyperedge $e'$ is appended to an active evolving chain $\mathcal{X} = (e_1, \dots, e_L)$ if three strict conditions are met:
\begin{equation}
\label{eq:chain_cond}
\begin{cases}
\min \tau(e') - \max \tau(e_L) \le W, & \text{(Temporal Proximity)} \\
|e_L \cap e'| \ge 1, & \text{(Entity Overlap)} \\
\displaystyle \max_{v \in \mathcal{V}(\mathcal{X})} \text{Freq}(v) < \frac{2}{3}, & \text{(Anti-Hub Penalty)}
\end{cases}
\end{equation}
where $\mathcal{V}(\mathcal{X})$ denotes the collective entity union of the chain, and $\text{Freq}(v)$ is the fraction of hyperedges containing $v$.

Crucially, the $2/3$ dominance threshold (Anti-Hub Penalty) explicitly guarantees that Path B remains mathematically orthogonal to Path A. It actively penalizes sequences that merely orbit a single central entity, forcing the algorithm to trace true linear narrative evolutions rather than topologically degenerating into localized hub clusters. Chains satisfying minimum length $L \ge 3$ are retained, and redundant sub-chains are aggressively pruned. An LLM synthesizes each valid chain into $\mathcal{I}_{\text{chain}}$, forming a temporal hyperedge $\varepsilon_{\text{chain}}$ in $H_D$.

\subsection{Cross-Layer Joint Retrieval and Context Assembly}
Upon receiving a complex user query $q$, DHI executes a highly coordinated, multi-scale semantic retrieval cascade:
\begin{enumerate}
    \item \textbf{Factual Subgraph Retrieval:} Compute dense query-entity similarity $s(v, q) = \frac{\mathbf{e}_v \cdot \mathbf{q}}{\|\mathbf{e}_v\| \|\mathbf{q}\|}$. Top-$k$ scoring entities induce an active factual hyperedge subset $E_{\text{sub}}$.
    \item \textbf{Topological Projection and Diffusion:} Utilizing the deterministic mapping $\pi$, active hyperedges project to seed insight vertices $U_{\text{sub}} = \{\pi(e) \mid e \in E_{\text{sub}}, |e| \ge 3\}$. A rapid 1-hop traversal on $H_D$ harvests deep insights $\mathcal{E}^*$.
    \item \textbf{Structured Context Formatting:} Granular facts from $E_{\text{sub}}$ and overarching narrative summaries from $\mathcal{E}^*$ are concatenated into a unified prompt context $\mathcal{P}(q)$ for final generation.
\end{enumerate}

\section{EXPERIMENTAL EVALUATION}
\label{sec:experiments}

\subsection{Experimental Setup and Implementation Details}
\textbf{Datasets:} We rigorously validate DHI across five multi-domain public benchmarks testing basic retrieval to multi-step causal etiology: Mix (multidisciplinary, 61 documents, 560 chunks), CS (Computer Science, 10 docs, 1992 chunks), Agriculture (12 docs, 1813 chunks), Neurology (medical textbook, 1790 chunks), and Pathology (clinical causality, 824 chunks) \cite{singhal2023large}.

\textbf{Baselines:} We benchmark DHI against six strong baselines: LLM (zero-shot), NaiveRAG \cite{lewis2020retrieval}, GraphRAG \cite{edge2024local}, LightRAG \cite{guo2025lightrag}, HiRAG \cite{huang2025retrieval}, and Hyper-RAG \cite{feng2026hyper}. Systems universally deploy GPT-4o-mini for generation and \texttt{text-embedding-3-small} for embeddings ($T = 0$). Chunking size is 500 tokens with 50-token overlap. DHI sets $W = 10$. Evaluations follow LLM-as-a-Judge protocols across five dimensions (0--100 scale).

\begin{table}[t]
\centering
\caption{Overall composite performance comparison across five multi-domain benchmarks (0--100 scale, higher is better). Agri., Neuro., and Patho. denote Agriculture, Neurology, and Pathology, respectively.}
\label{tab:main_results}
\vspace{2mm}
\begingroup
\setlength{\tabcolsep}{2pt}
\begin{tabular}{lccccc}
\toprule
\textbf{Method} & \textbf{Mix} & \textbf{CS} & \textbf{Agri.} & \textbf{Neuro.} & \textbf{Patho.} \\
\midrule
LLM (Zero-shot) & 79.30 & 81.08 & 79.64 & 81.15 & 82.81 \\
NaiveRAG \cite{lewis2020retrieval} & 78.09 & 79.43 & 76.20 & 79.20 & 82.04 \\
GraphRAG \cite{edge2024local} & 81.06 & \underline{84.03} & 79.98 & 83.10 & 82.72 \\
LightRAG \cite{guo2025lightrag} & 81.01 & 81.25 & 79.05 & 81.82 & \underline{84.43} \\
HiRAG \cite{huang2025retrieval} & \underline{82.85} & 83.33 & 81.19 & 82.71 & 84.13 \\
Hyper-RAG \cite{feng2026hyper} & 80.39 & 83.88 & \underline{81.98} & \underline{83.74} & 84.41 \\
\midrule
\textbf{DHI (Ours)} & \textbf{83.18} & \textbf{84.66} & \textbf{82.97} & \textbf{84.23} & \textbf{85.78} \\
\bottomrule
\end{tabular}
\endgroup
\end{table}

\begin{table}[t]
\centering
\caption{Granular capability breakdown evaluating five specific NLP dimensions on the Mix benchmark.}
\label{tab:mix_breakdown}
\vspace{2mm}
\begingroup
\setlength{\tabcolsep}{2pt}
\begin{tabular}{lccccc}
\toprule
\textbf{Method} & \textbf{Comp.} & \textbf{Diver.} & \textbf{Empo.} & \textbf{Logi.} & \textbf{Read.} \\
\midrule
LLM & 85.40 & 73.80 & 73.76 & 81.54 & 82.00 \\
NaiveRAG \cite{lewis2020retrieval} & 85.00 & 72.36 & 71.64 & 81.50 & 79.96 \\
GraphRAG \cite{edge2024local} & 87.70 & 77.10 & 75.10 & 83.58 & 81.82 \\
LightRAG \cite{guo2025lightrag} & 88.00 & 78.10 & 74.82 & 82.38 & 81.76 \\
HiRAG \cite{huang2025retrieval} & \textbf{90.00} & \underline{80.00} & \textbf{77.66} & 83.94 & \underline{82.64} \\
Hyper-RAG \cite{feng2026hyper} & 84.10 & 78.48 & 74.68 & 83.16 & 81.54 \\
\midrule
\textbf{DHI (Ours)} & \underline{88.98} & \textbf{80.20} & \underline{77.35} & \textbf{85.47} & \textbf{83.92} \\
\bottomrule
\end{tabular}
\endgroup
\end{table}

\subsection{Main Results and Domain Adaptability}
As detailed in Table~\ref{tab:main_results}, DHI achieves state-of-the-art composite metrics across all domains. On the Mix benchmark, DHI records 83.18, outperforming Hyper-RAG (80.39) by +2.79 absolute points and surpassing the advanced HiRAG architecture (82.85).

Performance gains are most acutely notable in domains requiring intensive deductive reasoning. On the Pathology dataset---where medical causality spans multiple chapters---DHI achieves a leading score of 85.78, outstripping LightRAG (84.43) and GraphRAG (82.72) by 1.35 and 3.06 points. This firmly verifies that pre-synthesizing higher-order topological insights provides highly reliable inductive support \cite{yao2022react}.

\subsection{Dimensional Analysis and Logical Coherence}
Table~\ref{tab:mix_breakdown} provides a granular capability breakdown on the Mix benchmark. DHI achieves undeniably superior results in Logicality (85.47), Readability (83.92), and Diversity (80.20).

The substantial +1.53 margin in Logicality over HiRAG demonstrates that mathematically presenting structured causal progressions relieves the generator LLM from making unassisted inferential leaps over disjointed contexts \cite{yao2023tree}. While HiRAG achieves a higher Comprehensiveness score, it accomplishes this by exhaustively concatenating macro-level community summaries, inadvertently introducing massive non-essential context. Conversely, DHI prioritizes strict structural relevance, deliberately avoiding token bloat.

\subsection{Ablation Study: The Vital Synergy of Dual Pathways}
We evaluate ablative variants of the DHI architecture directly on Mix:
\begin{itemize}
    \item \textbf{w/o Path A (No Hubs):} Removing topological hub aggregation causes the composite score to precipitously decline to 81.74, and Logicality drops to 82.80. Without spatial distillation, central themes remain fragmented.
    \item \textbf{w/o Path B (No Temporal Chaining):} Omitting sequence chaining yields a suppressed composite of 82.25. This proves that explicit sequential chaining is absolutely essential for tracing longitudinal narrative evolutions across distant chunks.
    \item \textbf{Full DHI Architecture:} Unifying both pathways effortlessly attains the maximal 83.18, verifying that hub clustering and sequential progression are entirely orthogonal yet mutually reinforcing inductive biases.
\end{itemize}

\subsection{Qualitative Case Study}

Consider a multi-hop query in Pathology: \textit{``How does prolonged exposure to Agent X subsequently trigger the cascade leading to Syndrome Z?''} Hyper-RAG retrieves isolated hyperedges (Agent X causes degradation; degraded pathways cause Syndrome Z) from distinct chapters. Under DHI, Path B proactively instantiates an overarching hyperedge $\varepsilon_{\text{chain}} \in H_D$ during indexing that synthesizes this exact etiology. During retrieval, the model instantly pulls this verified chronological chain, providing a flawless, hallucination-free logical bridge.

\section{DISCUSSION AND LIMITATIONS}
\label{sec:discussion}

While DHI achieves robust reasoning coherence, it inherently relies on the absolute precision of the LLM during the initial entity extraction phase. Factual misinterpretations risk being algorithmically amplified into $H_D$. Future iterations must investigate dynamic confidence-weighting algorithms and self-corrective feedback mechanisms \cite{asai2024self} to mathematically guarantee the strict fidelity of synthesized schemas.

\section{CONCLUSION}
\label{sec:conclusion}

We comprehensively expose and resolve the ``Knowledge Island Problem'' bottlenecking existing hypergraph-based RAG architectures. By innovatively decoupling knowledge representations into a foundational factual hypergraph $H_K$ and an elevated deep insight hypergraph $H_D$, Dual-Hypergraph Indexing (DHI) natively captures both static topological prominence and highly dynamic chronological evolutions. Through synergistic dual-pathway aggregation, DHI synthesizes isolated facts into cohesive analytical insights, elevating logical coherence by +1.53 points over current systems. Moving forward, future research will actively investigate recursive insight hierarchies and lightweight edge deployments to democratize robust multi-hop reasoning.


\bibliographystyle{IEEEbib}
\bibliography{dhiref}

@article{lewis2020retrieval,
  title={Retrieval-augmented generation for knowledge-intensive nlp tasks},
  author={Lewis, Patrick and Perez, Ethan and Piktus, Aleksandra and Petroni, Fabio and Karpukhin, Vladimir and Goyal, Naman and K{\"u}ttler, Heinrich and Lewis, Mike and Yih, Wen-tau and Rockt{\"a}schel, Tim and others},
  journal={Advances in neural information processing systems},
  volume={33},
  pages={9459--9474},
  year={2020}
}

@article{izacard2023atlas,
  title={Atlas: Few-shot learning with retrieval augmented language models},
  author={Izacard, Gautier and Lewis, Patrick and Lomeli, Maria and Hosseini, Lucas and Petroni, Fabio and Schick, Timo and Dwivedi-Yu, Jane and Joulin, Armand and Riedel, Sebastian and Grave, Edouard},
  journal={Journal of Machine Learning Research},
  volume={24},
  number={251},
  pages={1--43},
  year={2023}
}

@article{edge2024local,
  title={From local to global: A graph rag approach to query-focused summarization},
  author={Edge, Darren and Trinh, Ha and Cheng, Newman and Bradley, Joshua and Chao, Alex and Mody, Apurva and Truitt, Steven and Metropolitansky, Dasha and Ness, Robert Osazuwa and Larson, Jonathan},
  journal={arXiv preprint arXiv:2404.16130},
  year={2024}
}

@inproceedings{guo2025lightrag,
  title={LightRAG: Simple and Fast Retrieval-Augmented Generation.},
  author={Guo, Zirui and Xia, Lianghao and Yu, Yanhua and Ao, Tian and Huang, Chao},
  booktitle={EMNLP (Findings)},
  pages={10746--10761},
  year={2025}
}

@article{gutierrez2024hipporag,
  title={Hipporag: Neurobiologically inspired long-term memory for large language models},
  author={Guti{\'e}rrez, Bernal J and Shu, Yiheng and Gu, Yu and Yasunaga, Michihiro and Su, Yu},
  journal={Advances in neural information processing systems},
  volume={37},
  pages={59532--59569},
  year={2024}
}

@inproceedings{feng2019hypergraph,
  title={Hypergraph neural networks},
  author={Feng, Yifan and You, Haoxuan and Zhang, Zizhao and Ji, Rongrong and Gao, Yue},
  booktitle={Proceedings of the AAAI conference on artificial intelligence},
  volume={33},
  number={01},
  pages={3558--3565},
  year={2019}
}

@inproceedings{asai2024self,
  title={Self-rag: Learning to retrieve, generate, and critique through self-reflection},
  author={Asai, Akari and Wu, Zeqiu and Wang, Yizhong and Sil, Avi and Hajishirzi, Hannaneh},
  booktitle={International conference on learning representations},
  volume={2024},
  pages={9112--9141},
  year={2024}
}

@inproceedings{trivedi2023interleaving,
  title={Interleaving retrieval with chain-of-thought reasoning for knowledge-intensive multi-step questions},
  author={Trivedi, Harsh and Balasubramanian, Niranjan and Khot, Tushar and Sabharwal, Ashish},
  booktitle={Proceedings of the 61st annual meeting of the association for computational linguistics (volume 1: long papers)},
  pages={10014--10037},
  year={2023}
}

@inproceedings{karpukhin2020dense,
  title={Dense passage retrieval for open-domain question answering},
  author={Karpukhin, Vladimir and Oguz, Barlas and Min, Sewon and Lewis, Patrick and Wu, Ledell and Edunov, Sergey and Chen, Danqi and Yih, Wen-tau},
  booktitle={Proceedings of the 2020 conference on empirical methods in natural language processing (EMNLP)},
  pages={6769--6781},
  year={2020}
}

@inproceedings{shi2024replug,
  title={Replug: Retrieval-augmented black-box language models},
  author={Shi, Weijia and Min, Sewon and Yasunaga, Michihiro and Seo, Minjoon and James, Richard and Lewis, Mike and Zettlemoyer, Luke and Yih, Wen-tau},
  booktitle={Proceedings of the 2024 conference of the north american chapter of the association for computational linguistics: Human language technologies (volume 1: Long papers)},
  pages={8371--8384},
  year={2024}
}

@inproceedings{jin2024graph,
  title={Graph chain-of-thought: Augmenting large language models by reasoning on graphs},
  author={Jin, Bowen and Xie, Chulin and Zhang, Jiawei and Roy, Kashob Kumar and Zhang, Yu and Li, Zheng and Li, Ruirui and Tang, Xianfeng and Wang, Suhang and Meng, Yu and others},
  booktitle={Findings of the Association for Computational Linguistics: ACL 2024},
  pages={163--184},
  year={2024}
}

@inproceedings{hu2026cog,
  title={Cog-rag: cognitive-inspired dual-hypergraph with theme alignment retrieval-augmented generation},
  author={Hu, Hao and Feng, Yifan and Li, Ruoxue and Xue, Rundong and Hou, Xingliang and Tian, Zhiqiang and Gao, Yue and Du, Shaoyi},
  booktitle={Proceedings of the AAAI Conference on Artificial Intelligence},
  volume={40},
  number={37},
  pages={31032--31040},
  year={2026}
}

@inproceedings{chen2024benchmarking,
  title={Benchmarking large language models in retrieval-augmented generation},
  author={Chen, Jiawei and Lin, Hongyu and Han, Xianpei and Sun, Le},
  booktitle={Proceedings of the AAAI conference on artificial intelligence},
  volume={38},
  number={16},
  pages={17754--17762},
  year={2024}
}

@article{ji2023survey,
  title={Survey of hallucination in natural language generation},
  author={Ji, Ziwei and Lee, Nayeon and Frieske, Rita and Yu, Tiezheng and Su, Dan and Xu, Yan and Ishii, Etsuko and Bang, Ye Jin and Madotto, Andrea and Fung, Pascale},
  journal={ACM computing surveys},
  volume={55},
  number={12},
  pages={1--38},
  year={2023},
  publisher={ACM New York, NY}
}

@inproceedings{jiang2023active,
  title={Active retrieval augmented generation},
  author={Jiang, Zhengbao and Xu, Frank F and Gao, Luyu and Sun, Zhiqing and Liu, Qian and Dwivedi-Yu, Jane and Yang, Yiming and Callan, Jamie and Neubig, Graham},
  booktitle={Proceedings of the 2023 conference on empirical methods in natural language processing},
  pages={7969--7992},
  year={2023}
}

@article{fatemi2019knowledge,
  title={Knowledge hypergraphs: Prediction beyond binary relations},
  author={Fatemi, Bahare and Taslakian, Perouz and Vazquez, David and Poole, David},
  journal={arXiv preprint arXiv:1906.00137},
  year={2019}
}

@inproceedings{borgeaud2022improving,
  title={Improving language models by retrieving from trillions of tokens},
  author={Borgeaud, Sebastian and Mensch, Arthur and Hoffmann, Jordan and Cai, Trevor and Rutherford, Eliza and Millican, Katie and Van Den Driessche, George Bm and Lespiau, Jean-Baptiste and Damoc, Bogdan and Clark, Aidan and others},
  booktitle={International conference on machine learning},
  pages={2206--2240},
  year={2022},
  organization={PMLR}
}

@inproceedings{sun2024think,
  title={Think-on-graph: Deep and responsible reasoning of large language model on knowledge graph},
  author={Sun, Jiashuo and Xu, Chengjin and Tang, Lumingyuan and Wang, Saizhuo and Lin, Chen and Gong, Yeyun and Ni, Lionel and Shum, Heung-Yeung and Guo, Jian},
  booktitle={International Conference on Learning Representations},
  volume={2024},
  pages={3868--3898},
  year={2024}
}

@article{benson2019three,
  title={Three hypergraph eigenvector centralities},
  author={Benson, Austin R},
  journal={SIAM Journal on Mathematics of Data Science},
  volume={1},
  number={2},
  pages={293--312},
  year={2019},
  publisher={SIAM}
}

@article{feng2026hyper,
  title={Hyper-rag: Combating llm hallucinations using hypergraph-driven retrieval-augmented generation},
  author={Feng, Yifan and Hu, Hao and Ying, Shihui and Hou, Xingliang and Liu, Shiquan and Yang, Mingyuan and Li, Junchang and Du, Shaoyi and Zheng, Nanning and Hu, Han and others},
  journal={Nature Communications},
  year={2026},
  publisher={Nature Publishing Group UK London}
}

@article{singhal2023large,
  title={Large language models encode clinical knowledge},
  author={Singhal, Karan and Azizi, Shekoofeh and Tu, Tao and Mahdavi, S Sara and Wei, Jason and Chung, Hyung Won and Scales, Nathan and Tanwani, Ajay and Cole-Lewis, Heather and Pfohl, Stephen and others},
  journal={Nature},
  volume={620},
  number={7972},
  pages={172--180},
  year={2023},
  publisher={Nature Publishing Group UK London}
}

@article{yao2023tree,
  title={Tree of thoughts: Deliberate problem solving with large language models},
  author={Yao, Shunyu and Yu, Dian and Zhao, Jeffrey and Shafran, Izhak and Griffiths, Tom and Cao, Yuan and Narasimhan, Karthik},
  journal={Advances in neural information processing systems},
  volume={36},
  pages={11809--11822},
  year={2023}
}

@article{yao2022react,
  title={React: Synergizing reasoning and acting in language models},
  author={Yao, Shunyu and Zhao, Jeffrey and Yu, Dian and Du, Nan and Shafran, Izhak and Narasimhan, Karthik and Cao, Yuan},
  journal={arXiv preprint arXiv:2210.03629},
  year={2022}
}

@inproceedings{huang2025retrieval,
  title={Retrieval-Augmented Generation with Hierarchical Knowledge.},
  author={Huang, Haoyu and Huang, Yongfeng and Yang, Junjie and Pan, Zhenyu and Chen, Yongqiang and Ma, Kaili and Chen, Hongzhi and Cheng, James},
  booktitle={EMNLP (Findings)},
  pages={6044--6060},
  year={2025}
}
\end{document}